\documentclass[12pt]{article}
\usepackage{epsfig}
\usepackage{a4,isolatin1}
\usepackage{amsmath,amsfonts,latexsym, amssymb}

\newtheorem{satz}{Theorem}[section]
\newtheorem{defi}[satz]{Definition}

\newtheorem{bem}[satz]{Remark}
\newtheorem{lemma}[satz]{Lemma}

\newtheorem{ob}[satz]{Observation}

\newtheorem{propo}[satz]{Proposition}

\newcommand{\mcal}{\mathcal}

\newcommand{\tit}{\textit}

\newcommand{\C}{\mathbb{C}}

\newcommand{\Z}{\mathbb{Z}}

\newcommand{\dast}{d^{\ast}}

\begin{document}
\thispagestyle{empty}
\begin{center}
\vspace*{1.0cm}

{\LARGE{\bf Supersymmetry on Graphs and Networks}} 

\vskip 1.5cm

{\large {\bf Manfred Requardt }} 

\vskip 0.5 cm 

Institut f\"ur Theoretische Physik \\ 
Universit\"at G\"ottingen \\ 
Friedrich-Hund-Platz 1 \\ 
37077 G\"ottingen \quad Germany\\
(E-mail: requardt@theorie.physik.uni-goettingen.de)

\end{center}

\vspace{1 cm}

\begin{abstract}
We show that graphs, networks and other related discrete model systems 
carry a natural supersymmetric structure, which, apart from its
conceptual importance as to possible physical applications, allows to
derive a series of spectral properties for a class of graph operators
which typically encode relevant graph characteristics.
\end{abstract} \newpage
\setcounter{page}{1}
\section{Introduction}
In the following we show that graphs and networks support in a very
natural way a supersymmetric structure. Our starting point is a class
of geometrically relevant \tit{graph-operators} acting on the direct
sum, $\mcal{H}$, of the \tit{vertex-Hilbert space}, $\mcal{H}_0$ and
the \tit{edge-Hilbert space}, $\mcal{H}_1$. As to these technical
prerequisites cf. the analysis being made in \cite{graphs} and further
references given there. 

Of particular importance is a \tit{graph-Dirac operator}, $D$,
introduced there, which maps $\mcal{H}_0$ into $\mcal{H}_1$ and vice
versa. Furthermore we have a natural \tit{Laplace operator} on
$\mcal{H}$ which together with $D$ and another \tit{supercharge} forms
a closed \tit{($N=2$)-superalgebra}. This abstract susy-structure, if
concretely represented by our graph-operators, allows, among other
things, to derive a series of useful spectral properties of these
operators.

The natural existence of this susy structure on graphs (and related
models) may also be of some relevance in a wider context. We have been
promoting a discrete network approach to quantum space-time physics in
recent years (see for example \cite{Requ1} or \cite{Requ2} and further
references given there) which is assumed to underly our more
macroscopic continuum physics on the Planck-scale. Our present
analysis shows that these discrete model systems, perhaps contrary to
naive wisdom, are in fact quite rich as to their structural properties.

We conclude this introduction with the remark that such technical
properties of graphs have also been mentioned elsewhere in a however
different context and using a different formalism (cf.
\cite{Ogurisu}).  So we should add the remark that there may exist
papers in other fields of research being related to our work but which
we presently are not aware of.

\section{Some Concepts from Algebraic Graph Theory}
To set the stage, we briefly introduce some concepts employed in
\tit{algebraic graph theory} but in a form particularly adapted to our
own framework (see also \cite{graphs}; as to the wider context cf.
\cite{Biggs} to \cite{Woess} and further references given there). The
graph, $G$, consists of a countable set of vertices (or nodes), $V$,
with $x_i$ the \tit{labelled} vertices, and an \tit{edge set}, $E$,
the \tit{directed} edges, pointing from $x_i$ to $x_j$, denoted by
$d_{ij}$. We assume (for convenience) the graph to be \tit{simple},
that is, an edge, $d_{ij}$, can be associated with the ordered pair of
vertices, $(x_i,x_j)$ with $i\neq j$.

So-called unoriented graphs with the edges consisting of unordered
pairs, $\{x_i,x_j\}$, can be subsumed in the above framework in two
different ways, each of which having a certain advantage of its own.
On the one hand, we can give the graph an arbitrary \tit{orientation},
that is we associate to each undirected edge,$\{x_i,x_j\}$, one of the
two possible choices, $(x_i,x_j)$ or $(x_j,x_i)$. It turns out that
most of the concepts and calculations do not depend on the particular
choice (see below). On the other hand, we can associate an unoriented
but orientable graph with a directed graph so that each edge occurs
twice, that is if $d_{ij}$ belongs to $E$, $d_{ji}$ also belongs to
$E$.

For reasons of simplicity we assume our graph to be \tit{locally
  finite}, that is, each vertex is only incident with a finite number
of edges. A slightly stronger assumption is it to assume this
\tit{vertex degree} to be globally finite over the vertex set. For a
directed graph we then have ingoing edges and outgoing edges relative
to a given node with the respective vertex degrees, $v_i^{in}$ and    
$v_i^{out}$ and the total degree $v_i=v_i^{in}+v_i^{out}$.

For such a graph we can introduce two Hilbert spaces, a vertex Hilbert
space, $\mcal{H}_0$, and an edge Hilbert space, $\mcal{H}_1$, with
orthonormal bases the set of vertices, $x_i$, and the set of directed
edges, $d_{ij}$. This means, we introduce a formal scalar product on
$\mcal{H}_0,\mcal{H}_1$ respectively with
\begin{equation}(x_i,x_j)=\delta_{ij}\quad,\quad
  (d_{ij},d_{lm})=\delta_{il}\delta_{jm}   \end{equation}
and with vectors being the formal sums
\begin{equation}f=\sum_1^{\infty}f_ix_i\quad,\quad
  g=\sum_{i,j=1}^{\infty}g_{ij}d_{ij}\;\text{with}\; f_i,g_{ij}\in\C
\end{equation} 
with $\sum |f_i|^2<\infty$ and $\sum |g_{ik}|^2<\infty$.\\[0.3cm]
Remark: We treat the vertices and edges as abstract basis elements (in
a way similar to the \tit{group algebra} of a group). One can of
course consider the abstract vectors equally well as discrete
functions over the vertex- or edge-set, respectively and the basis
vectors as elementary indicator functions. \vspace{0.3cm}

If we deal with an undirected but orientable graph, employing the
second variant introduced above, we found it
convenient (cf. \cite{graphs}) to introduce the superposition
\begin{equation}b_{ij}:=d_{ij}-d_{ji}=-b_{ji}    \end{equation}
and relate it to an undirected but orientable edge.
\section{Operators on Graphs}
In \cite{graphs} we introduced two operators, interpolating between
$\mcal{H}_0$ and $\mcal{H}_1$. We define them on the basis vectors:
\begin{equation}d(x_i):=\sum_k d_{ki}-\sum_{k'}d_{ik'}
\end{equation}
with the first sum running over the ingoing edges relative to $x_i$,
the second sum running over the outgoing edges. In the case of a
symmetric (or undirected graph; version two) we have
\begin{equation}d(x_i):=\sum_k \left(d_{ki}-d_{ik}\right)=\sum_k
  b_{ki}    \end{equation} 
This operator is closely related to a sort of non-commutative discrete
differential calculus on graphs as we have
\begin{equation}df=\sum_{i,k}(f_k-f_i)d_{ik}      \end{equation}
A simple calculation shows that the adjoint, $d^*$, acts on the basis
vectors of $\mcal{H}_1$ as follows:
\begin{equation}d^*(d_{ik})=x_k-x_i      \end{equation}

In algebraic graph theory (finite graphs) the so-called \tit{incidence
matrix}, $B$, is introduced, having the entry $1$ if vertex $x_i$ is
the positive end of a certain (ingoing) edge, and having a $-1$ if it is the
negative end (outgoing edge) (see for example \cite{Godsil}). This
matrix corresponds to our operator $d^*$.

Another important operator is the \tit{adjacency matrix}, $A,$ being a
map from $\mcal{H}_0$ to $\mcal{H}_0$ and having (in ordinary graph theory of
(un)oriented graphs) a $+1$ at entry $(i,j)$ if $x_i$ and $x_j$ are
connected by an edge. This matrix is a symmetric operator,
$a_{ij}=a_{ji}$. In our more general context (which includes however
the ordinary situation as a special case) of directed graphs one can
introduce the \tit{in-adjacency} matrix, $A^{in}$, and the
\tit{out-adjacency} matrix, $A^{out}$, with $A=A^{in}+A^{out}$. In our
(operator)-notation they are given by
\begin{equation}A\,x_i=\sum_{k\sim i}\epsilon_{ki}\,x_k\;,\;A^{in}\,x_i=\sum_{k\to
    i}x_k\;,\;A^{out}\,x_i=\sum_{i\to k'}x_{k'}      \end{equation}
with $\sim$ designating the unordered pair $\{x_i,x_k\}$, $k\to i$ the
ordered pair $(k,i)$ and $\epsilon_{k,i}$ is either one or two depending
on the two possible cases of one directed edge between node $x_i$ and
node $x_k$ or two directed edges, pointing in opposite directions.

These operators can be built up from more elementary operators (cf.
\cite{graphs}).
\begin{equation}d_1\,x_i=\sum_k d_{ki}\quad,\quad d_2\,x_i=\sum_{k'} d_{ik'}      \end{equation}
\begin{equation}d_1^*\;d_{ik}=x_k\quad,\quad d_2^*\,d_{ik}=x_i      \end{equation}
so that 
\begin{equation}d=d_1-d_2\quad,\quad d^*=d_1^*-d_2^*      \end{equation}
\begin{equation}d_1^*d_1\,x_i=v_i^{in}\cdot x_i\quad,\quad
  d_2^*d_2\,x_i=v_i^{out}\cdot x_i     \end{equation}
\begin{equation}d_1^*d_2\,x_i=\sum_{i\to k'}x_{k'}\quad,\quad
  d_2^*d_1\,x_i=\sum_{k\to i}x_k      \end{equation}
where $v_i^{in}\,,\,v_i^{out}$ is the in-, out degree of vertex $x_i$ respectively.
We hence have
\begin{lemma}The in-, out-vertex degree matrices read
\begin{equation}V^{in}=d_1^*d_1\quad,\quad V^{out}=d_2^*d_2     \end{equation}
The in-, out-adjacency matrices read
\begin{equation}A^{in}=d_2^*d_1\quad,\quad A^{out}=d_1^*d_2     \end{equation}
$A=A^{in}+A^{out}$ is symmetric.
\end{lemma}
\begin{propo}The so-called graph Laplacian is the following positive
  operator
\begin{equation}    -\Delta:=d^*d=\left(V^{in}+V^{out}\right)-\left(A^{in}+A^{out}\right)=V-A \end{equation}
\end{propo}

The reason to call this operator a Laplacian stems from the
observation that it acts like a second order partial difference
operator on functions of $H_0$.
\begin{equation}-\Delta\,f=\sum_i
  f_i\left(v^{in}_ix_i+v^{out}_ix_i-\sum_{k\to i}x_k-\sum_{i\to
      k}x_k\right)\end{equation} 
and after a simple relabelling of indices
\begin{multline}-\Delta\,f=-\sum_i\left(\sum_{k\to i}f_k+\sum_{i\to
      k}f_k-v^{in}_if_i-v^{out}_if_i\right)x_i\\
=-\sum_i\left(\sum_{k\to
    i}(f_k-f_i)+\sum_{i\to k}(f_k-f_i)\right)x_i=-\sum_i\left(\sum_{k\sim
    i}\epsilon_{ki}(f_k-f_i)\right)x_i\end{multline}
which reduces to the ordinary expression in the undirected case.

Forming now the direct some $\mcal{H}:=\mcal{H}_0\oplus\mcal{H}_1$, we
can introduce yet another important graph operator which closely
entangles geometric and functional analytic properties of graphs (and
similar structures); see \cite{graphs}.
\begin{defi}We define the graph Dirac operator as follows
\begin{equation}D:\,\mcal{H}\to\mcal{H}\;\text{with}\, D:=\left( \begin{array}{cc}0
      & d^{\ast}\\d & 0 \end{array} \right)\quad,\quad  H=\left(
    \begin{array}{c}\mcal{H}_0 \\ \mcal{H}_1 \end{array} \right)  \end{equation}
\end{defi}
\begin{ob}\begin{equation}D^2=D\,D=\left( \begin{array}{cc}\dast d & 0\\0 & d\dast \end{array}
\right)\end{equation}
with $d^*d=-\Delta$.
\end{ob}

The action of $dd^*$ on a basis vector $d_{ik}$ reads
\begin{equation}dd^*\,d_{ik}=d\,(x_k-x_i)=\sum_{k'}d_{k'k}-\sum_{k''}d_{kk''}-\sum_l   d_{li}+\sum_{l'}d_{il'} \end{equation}   
which, after some relabelling and introduction of the Kronecker delta
function can be written as
\begin{equation}dd^*\,d_{ik}=\sum_{m,j}\left(d_{mj}\delta_{jk}-d_{jm}\delta_{jk}-d_{mj}\delta_{ij}+d_{jm}\delta_{ij}\right)     \end{equation}
For a function $g=\sum g_{ik}d_{ik}$ we hence get
\begin{equation}dd^*\,g=\sum_{l,m}\left(\sum_i
    g_{im}-g_{mi}-g_{il}+g_{li}\right)d_{lm}     \end{equation}

In the simple case of the one-dimensional directed lattice, $\Z_1$,
with directed edges only pointing from $x_i\to x_{i+1}$, we get a
transparent expression:
\begin{equation}dd^*\,\left(\sum_i g_{i,i+1}d_{i,i+1}\right)=-\sum_i\left(g_{i+1,i+2}+g_{i-1,i}-2g_{i,i+1}\right)d_{i,i+1}     \end{equation}
i.e., it represents also a discrete second derivative operator on the
level of edges. For the directed lattice, $\Z_n$, it can be related to
what is called the \tit{vector Laplacian} in the continuum
(cf. \cite{Bronstein}. p.575). 

There is a pendant in the calculus of differential forms on general
Riemannian manifolds where, with the help of the \tit{Hodge-star
  operation}, we can construct a dual, $\delta$, to the ordinary
exterior derivative. The generalized Laplacian then reads
\begin{equation}-\Delta=\delta d+d\delta    \end{equation}
with $\delta$ (modulo certain combinatorial prefactors) corresponding to
our $d^*$ (see for example \cite{Felsager} or \cite{Hodge}).
\section{Susy on Graphs}
We introduce the following simple model of a supersymmetric algebra,
consisting of two hermitean charges, $Q_1\,,\,Q_2$, and a Hamiltonian 
\begin{align}
H_S=Q_1^2=Q_2^2\quad &\text{hence}\quad [H_S,Q_{1,2}]=0\\ 
\{Q_1,Q_2\}=0\quad &\text{hence}\quad\{Q_i,Q_j\}=2H_S\cdot \delta_{ij}\end{align}   

Defining $Q_+,Q_-$ as
\begin{equation}Q_+:=2^{-1}(Q_1+iQ_2)\quad,\quad Q_-:=2^{-1}(Q_1-iQ_2)    \end{equation}
implying
\begin{equation}Q_1=Q_++Q_-\quad,\quad Q_2=-i(Q_+-Q_-)    \end{equation}
we get
\begin{equation}Q_+^2=Q_-^2=0\quad,\quad H_S=\{Q_+,Q_-\}\quad,\quad [H_S,Q_{\pm}]=0   \end{equation}
That is, these three generators create a closed (N=2) susy-algebra.

We now make the following correspondence with our graph operators:
\begin{equation}H_S=D^2=D\,D=\left( \begin{array}{cc}\dast d & 0\\0 & d\dast \end{array}\right)    \end{equation}
\begin{equation}Q_+=\left( \begin{array}{cc}0 & 0\\d & 0
    \end{array}\right)\quad,\quad Q_-= \left( \begin{array}{cc}0 & d^*\\0 & 0
    \end{array}\right)                             \end{equation}
yielding
\begin{equation}Q_1= D=\left( \begin{array}{cc}0
      & d^{\ast}\\d & 0 \end{array} \right)\quad,\quad Q_2=\left( \begin{array}{cc}0
      &id^{\ast}\\-id & 0 \end{array} \right)   \end{equation}
In other words, the charge $Q_1$ is our original Dirac operator.
We can now check all the above abstract (anti) commutation relations
and find that they are fulfilled by our representation in form of
graph operators.
We see that $Q_+$ is essentially a map from the subspace $\mcal{H}_0$ to the subspace
$\mcal{H}_1$ and $Q_-$ from $\mcal{H}_1$ to $\mcal{H}_0$ while $Q_+$ vanishes on $\mcal{H}_1$,
$Q_-$ on $\mcal{H}_0$. Therefore we can tentatively associate $\mcal{H}_0$ with the
\tit{bosonic} and $\mcal{H}_1$ with the \tit{fermionic} subspace.
Furthermore there exists a natural \tit{grading operator} on $\mcal{H}$, given by
\begin{equation}\chi:=\left(\begin{array}{cc}1 & 0\\0 & -1
    \end{array}\right)    \end{equation}
We have $Q_2=iQ_1\cdot\chi$. Both $Q_1$ and $Q_2$ anticommute with
$\chi$ and are therefore called supercharges of an \tit{abstract}
supersymmetric quantum mechanics (cf. \cite{Kalka}, sect. 8.3 or
\cite{Thaller}, \cite{Jaffe} respectively). We note that many of the
susy properties follow already from the existence of this abstract
structure.
\begin{bem}Note that $\chi$ is both selfadjoint and unitary with
\begin{equation}\chi=\chi^*=\chi^{-1}\;,\;\chi^2=1    \end{equation}
The projectors on $\mcal{H}_0$ and $\mcal{H}_1$ are given by 
\begin{equation}P_0=1/2(1+\chi)\;,\;P_1=1/2(1-\chi)    \end{equation}
respectively, as for ordinary continuum Dirac operators.
\end{bem}

\section{Some Graph-Spectral Properties following from Susy}
We indicated already in \cite{graphs} (see also \cite{Biggs}) that a
variety of spectral properties on graphs are encoded in our Laplace or
Dirac operator.  For a finite, connected (for reasons of simplicity
only) graph we have for example
\begin{ob}\label{finite}The following operator kernel, range
  properties hold.
\begin{equation}Dim\left(Rg\,d^*\right)=n-1\;,\;Dim\left(Ker\,d^*\right)=Dim(Rg\,d)^{\bot})=\sum
  v_i^{in}-(n-1)\geq 0\end{equation} 
\begin{equation}Dim(Rg\,d)=Dim(Ker\,d^*)^{\bot})= n-1   \end{equation}
that is, $Rg\, d^*$ and $Rg\,d$ have the same dimension, $(n-1)$, and
the dimension of $Ker\,d$ is one (\,$n$ designates the order of the
graph i.e.the number of vertices). Note that $\sum v_i^{in}=\sum
v_i^{out}=\sum v_i/2=\#(edges)$ is the dimension of $\mcal{H}_1$ as
each (directed) edge occurs as an ingoing edge for exactly one node.
\end{ob}
To give an idea how these results can be proved, take for example the
first statement. $Ker\,d$ is spanned by the vector $\sum x_i$ as
each edge, occurring as an in-edge for, say, $x_i$
occurs as an out-edge for some other $x_j$, hence $d\left(\sum
  x_i\right)$ vanishes. With $Rg\,A^*=(Ker\,A)^{\bot}$ the result
follows and the other results follow from simple vector space
mathematics and properties of the adjoint (note for example that for
finite dimensions $Dim(Ker\,d)+Dim(Rg\,d)=n)$).

For general infinite graphs our susy structure allows to infer more
interesting spectral results. Note that, to keep matters simple, we
restricted ourselves to graphs with globally bounded node degree.
Therefore all our operators are bounded and there are hence no operator
domain problems. In the following the \tit{polar decomposition} of an
operator turns out to be useful. With A a closed operator from a Hilbert
space, $X_1$, to a Hilbert space, $X_2$ we have the following
representation
\begin{equation}A=S\cdot |A|   \end{equation}
with $|A|= (A^*A)^{1/2}$ a positive operator from $X_1$ to $X_1$, $S$
a partial isometry, mapping $Rg\,|A|$ isometrically onto $Rg\, A$ (see
for example \cite{Weidmann}).

We write $d$ as $d=S|d|$ with $|d|=(d^*d)^{1/2}$. We hence have
$d^*=|d|S^*$ and with $dd^*=|d^*|^2=S|d|^2S^*$ and consequently
$|d^*|=S|d|S^*$ (the square root):
\begin{equation}d=S|d|=|d^*|S\quad,\quad d^*=|d|S^*=S^*|d^*|   \end{equation}
For our $Q_1:=D$ we have
\begin{lemma}The polar decomposition of $Q_1=D$ is
\begin{equation}Q_1=\begin{pmatrix}0 & d^* \\d & 0
  \end{pmatrix}=\begin{pmatrix}0 & S^*|d^*| \\S|d| & 0
  \end{pmatrix}=\begin{pmatrix}0 & S^* \\S & 0 \end{pmatrix}\cdot\begin{pmatrix}|d| & 0 \\0 & |d^*| \end{pmatrix}    \end{equation}
\end{lemma}
We furthermore have
\begin{lemma}\label{Ker}
\begin{equation}Ker\,Q_i=Ker\,d\oplus Ker\,d^*=Ker\,Q_i^2    \end{equation}
with corresponding formulas holding for the respective orthogonal complements.
For a finite connected graph we hence have (by observation \ref{finite})
\begin{equation}Dim(Ker\,Q_{1,2})=\sum v_i^{in}-(n-1)+1=\#(edges)-(n-2)  \end{equation}
\end{lemma}
For the proof note that for e.g. $A$ bounded $Ker(A^*A)=Ker(A)$ as $Ker(A^*)=Rg(A)^{\bot}$.
\begin{propo}On $(Ker\,d)^{\bot}\oplus (Ker\,d^*)^{\bot}$ we have
\begin{equation}\begin{pmatrix}d^*d & 0 \\0 & dd^* \end{pmatrix}= \begin{pmatrix}S^*dd^*S & 0 \\0 & Sd^*dS \end{pmatrix}   \end{equation}
That is, $d^*d$ on $Rg\,d^*$ is unitarily equivalent to $dd^*$ on
$Rg\,d$. This is a generalisation to infinite dimensional spaces of a
previous result concerning the dimension of the respective subspaces. It
follows that the spectra of $d^*d$ and $dd^*$ coincide away from
zero! We have
\begin{equation}d^*d\,f=E\,f\;\Rightarrow\;dd^*\,S\,f=E\,S\,f   \end{equation}
\end{propo}
Therefore the eigenvalue spectrum of $H_S$ away from zero is at least
twofold degenerate.
\begin{equation}H_S\,(f,g)^T=E\,(f,g)^T\;\Rightarrow\;d^*d\,f=E\,f\;\text{and}\;dd^*\,g=E\,g  \end{equation}
hence, $(f,0)^T$ and $(0,g)^T$ are eigenvectors of $H_S$ to
the same eigenvalue and are at the same time eigenvectors of $\chi$ to
the eigenvalues $\pm 1$.

Furthermore, as $H_S=Q_1^2=Q_2^2$ and $\{Q_1,Q_2\}=0$, certain
combinations of the above eigenvectors yield common eigenvectors of
the pairs $H_S,Q_i$.
\begin{propo}The eigenvalues away from zero of $H_S=Q_i^2$ are at least twofold
  degenerate. With $E>0$ being an eigenvalue of $H_S=Q_1^2=Q_2^2$ with
  $(f,g)^T$ the corresponding eigenvector of $Q_1$, the eigenvalue of
  $Q_1$ is $+E^{1/2}$ or $-E^{1/2}$ and, due to the anticommutation
  relation between $Q_1$ and $Q_2$, $Q_2\,(f,g)^T$ is another
  eigenvector to $Q_1$ with eigenvalue $-E^{1/2}$ or $+E^{1/2}$. We
  thus see that both $Q_1$ and $Q_2$ have a symmetric eigenvalue
  spectrum.
\end{propo} 
We can make the result a little bit more explicit by making canonical
choices. 
\begin{ob}With $(f,g)^T$ an eigenvector of $Q_1$ to eigenvalue
  $\lambda$ (and hence $(f,0)^T\,,\,(0,g)^T$ ``pure'' eigenvectors of
  $H_S$ to eigenvalue $\lambda^2$), $(f,-g)^T$ is an eigenvector to
  eigenvalue $-\lambda$. Correspondingly, a straightforward
  calculation shows that $(if,g)^T$ is an eigenvector to $Q_2$ with
  eigenvalue $\lambda$ and $(if,-g)^T$ the eigenvector to eigenvalue
  $-\lambda$. All these vectors are eigenvectors of $H_S$ to
  eigenvalue $E=\lambda^2$. However, only the pairs belonging to $Q_1$
  or $Q_2$, respectively, are linearly independent. We see that
  $Q_{1,2}$ necessarily mix the pure bosonic and fermionic
  eigenstates of $H_S$ (see also sect. 2.3.3 of \cite{Kalka}).
\end{ob}

Boilt down to the two operators $d^*d$ and $dd^*$ we have that with
$f$ being an eigenvector of $d^*d$, $d\,f$ is an eigenvector of $dd^*$
to the same eigenvalue with a corresponding result for $dd^*\,,\,g$
and $d^*g$ ($d\,d^*d\,f=dd^*\,d\,f=E\,d\,f$). Hence, all eigenvectors
of $dd^*$ are of the form $g=d\,f$, $f$ an eigenvector of $d^*d$, both
belonging to the same eigenvalue (see also \cite{Thaller} sect.
5.2.3).  Furthermore, the eigenvectors, $(f,g)^T$ of, for example,
$Q_1$ are characterized by the following symmetry property:
\begin{equation}d^*\,g=\lambda\,f\quad,\quad d\,f=\lambda\,g
\end{equation}
We conclude from our preceding findings that the susy structure
relates the spectral properties of $d,d^*,d^*d,dd^*$ to each other. On
the other hand it seems to say (at least as far as we can see) not
much about the spectrum of, for example, $d^*d$ as such.
\section{The Zero Eigenspace}
In supersymmetric quantum mechanics the zero eigenspace of $H_S$ is
particularly interesting and is associated with the notion of
\tit{supersymmetry breaking} (see for example \cite{Kalka}). Our above
lemma \ref{Ker} makes an explicit statement about the dimension for a
finite connected graph. We see that the eigenspace is in many cases
highly degenerate and always has a dimension bigger than zero. This
can be seen as follows. Each connected graph contains a spanning tree
(see e.g. \cite{Bollo}). A finite tree has $(n-1)$ edges (see
below). Hence each connected graph contains at least $(n-1)$ edges so
that
\begin{equation}Dim(Ker H_S)\geq 1     \end{equation}
In case of a finite tree we have the following.

A finite tree of order $n$ has $(n-1)$ edges. This can most easily be
seen by choosing a base vertex, $x_0$, and then starting from the
outer vertices. Each edge corresponds to exactly one vertex, ending with
the base vertex, which does not correspond to an edge.
\begin{lemma}For a finite tree the preceding formula reduces to
\begin{equation}Ker\,d^*=(n-1)-(n-1)=0     \end{equation}
We infer that $Dim(Ker\,H_S)=1$.
\end{lemma}

The situation is more complicated for infinite connected graphs. In
this case the kernel of $d$ is zero. The formula
\begin{equation}0=d\,f=\sum_{ik}(f_i-f_k)d_{ik}      \end{equation}
implies $f_i=f_k$ for all pairs $(i,k)$ occurring in the sum. The only
normalisable vector in the infinite case has $f_i=0$ for all $i$.

On the other hand, for an oriented or (more generally) directed graph
each cycle lies in the zero eigenspace of $d^*$. More specifically,
denoting the cycle by the edge sequence $x_{i_1},\ldots,x_{i_l}$ with
$x_{i_l}$ linked to $x_{i_1}$, we choose the following vector in
$\mcal{H}_1$:
\begin{equation}g:=\sum_{\nu}[d_{i_{\nu},i{\nu+1}}]     \end{equation}
with $[d_{i_{\nu},i{\nu+1}}]$ denoting either $d_{i_{\nu},i{\nu+1}}$
if the edge is pointing from $x_{i_{\nu}}$ to $x_{i_{\nu+1}}$ or
$-d_{i_{\nu+1},i{\nu}}$ if it points in the opposite direction.
Applying $d^*$ to this vector yields the nullvector in $\mcal{H}_0$.
Furthermore we state without proof that (genuinely) different cycles
are linearly independent and span the kernel of $d^*$. This can be shown
by exploiting the existence of a \tit{spanning tree} (see, for example
\cite{Bollo}, p.53).  We then have
\begin{ob}For an infinite connected directed graph the zero eigenspace
  of $Q_1$, $Q_2$ or $H_S$ is purely fermionic and consists of the
  cycle space.
\end{ob}

This implies that the susy-Hamiltonian for an infinite tree has no
non-trivial zero eigenvectors. We provide a separate proof for this
statement as it employs a possibly useful technical property. We again
pick a base vertex, $x_0$, and construct the spheres
\begin{equation}\Gamma_l(x_0):=\{x_i\,,\,d(x_0,x_i)= l\}       \end{equation}
with $d$ denoting the canonical \tit{graph metric}. We can infer that
for a tree all the $(x_i-x_k)$ so that $x_i,x_k$ are linked by an edge
are linearly independent in $\mcal{H}_0$ and, together with $x_0$ span
$\mcal{H}_0$. We then have that
\begin{equation}0=d^*\,g=\sum g_{ik}(x_i-x_k)       \end{equation}
implies $g_{ik}=0$ for all occurring pairs.
\begin{bem}This fact can also be exploited for general graphs by using
  a spanning tree.
\end{bem}

\end{document}